**One-dimensional electronic solitons of graphene in an electromagnetic field**

Ji Luo[a]

*Department of Physics and Institute for Functional Nanomaterials,*

*University of Puerto Rico at Mayagüez, Mayagüez, PR 00681, USA*

Electronic energy-eigen-states of graphene in an orthogonal electromagnetic field with relative magnitude $\beta = E/v_F B \geq 1$ or in a pure electric field are obtained by a differential-equation method. Gaussian wave packets of probability density are constructed and found to transform form two-dimensional solitons into one-dimension ones at $\beta = 1$, when $\beta$ increases form $\beta < 1$ into $\beta > 1$. The maximum number of one-dimensional solitons of a finite graphene in the field is used to determine the electronic energy levels. These energy levels and the velocity of solitons lead to nonlinear dependence of the current on the Hall voltage for the field with $\beta > 1$ and nonlinear current-voltage curves for the pure electric field including the step-like ones at low temperatures, according to the ballistic transport of solitons.

---

[a]Electronic mail: ji.luo@upr.edu

The electronic energy-eigen-states of graphene in an orthogonal electromagnetic field $\vec{E}$ and $\vec{B}$ are the basis of its many properties including quantum Hall effect. At low excitation, graphene's electrons are described by two envelope functions which are determined by a Dirac-like Hamiltonian (DH), according to tight-binding $\vec{k} \cdot \vec{p}$ approximation.[1-4] Eigen-states of the DH have different features according to different relative magnitude of the field $\beta = E/v_F B$ with $v_F \approx 10^6$ ms$^{-1}$ the Fermi velocity. For a field with $\beta < 1$, eigen-states have been obtained analytically by using Lorentz transform[5] or algebraic properties of the DH.[6] It was predicted that when $\beta \to 1$, Landau-level (LL) collapse occurs, indicating the termination of separate LLs and corresponding bound states. For a field with $\beta \geq 1$ or a pure electric field, however, eigen-states are still unknown.

Initially graphene's Hall effect was demonstrated in linear response region with a strong magnetic field so that $\beta \ll 1$.[7-9] Later the effect in a field with $\beta \geq 1$ has also been investigated experimentally, and quasi-classical approach was adopted for theoretical explanation.[10] Eigen-states for $\beta \geq 1$ are indispensible for obtaining graphene's electronic properties in a weak magnetic field more exactly. In addition, when the magnetic field was not involved, most electronic transport investigations were based on eigen-states of field-free graphene, although the role of electric field was crucial.[11-16] Approaches adopting eigen-states of graphene in an electric field will be interesting if these states can be obtained.

In this work, eigen-states of graphene in a field with $\beta < 1$ are rederived by using a differential-equation method and then, eigen-states in a field with $\beta \geq 1$ or in a pure electric field are obtained by applying the same method. Gaussian wave packets of electronic probability density are constructed from the eigen-states. These wave packets propagate at fixed velocities as two-dimensional (2D) solitons for $\beta < 1$ and one-dimensional (1D) ones for $\beta \geq 1$ or a pure electric field. For $\beta < 1$, the equivalence of the number of states for each LL and the maximum number of solitions in finite graphene is proven, and this equivalence is generalized to obtain energy levels for $\beta \geq 1$ or a pure electric field. The ballistic transport of the 1D solitons is then used to calculate nonlinear dependence of the current on the Hall voltage for $\beta \geq 1$ and nonlinear current-voltage curves for a pure electric field.

Suppose a uniform parallel electric field $\vec{E}$ and a uniform perpendicular magnetic field $\vec{B}$ are applied to the graphene. In a general case, one set of carbon-carbon bonds has an arbitrary angle $\alpha$ with respect to the electric field. The graphene is taken as the $xy$-plane with $x$-axis parallel to the electric field. Unit vectors of axes are denoted by $\vec{x}$ and $\vec{y}$. Two kinds of Dirac points $\vec{k}_F = (4\pi/3\sqrt{3}a_0)[\cos(\alpha + \tau\pi/6)\vec{x} + \sin(\alpha + \tau\pi/6)\vec{y}]$ are distinguished by $\tau = \pm 1$. The electromagnetic field has scalar potential $\varphi(x,y) = -Ex$ and vector potential $\vec{A}(x,y) = Bx\vec{y}$. With these arrangements the DH becomes

$$\hat{H} = v_F \vec{\sigma} \cdot (-i\hbar\nabla + eBx\vec{y}) + eExI_2, \tag{1}$$

where $\nabla = \vec{x}\partial/\partial x + \vec{y}\partial/\partial y$, $I_2$ is the $2 \times 2$ unit matrix, and

$$\vec{\sigma} = \begin{pmatrix} 0 & e^{i\tau\alpha} \\ e^{-i\tau\alpha} & 0 \end{pmatrix} \vec{x} + \begin{pmatrix} 0 & -i\tau e^{i\tau\alpha} \\ i\tau e^{-i\tau\alpha} & 0 \end{pmatrix} \vec{y}. \quad (2)$$

To solve the eigen-equation $\hat{H}(\psi_1 \ \psi_2)^T = \varepsilon(\psi_1 \ \psi_2)^T$ with $\varepsilon$ the eigen-energy and $\psi_{1,2}$ the envelope functions, one separates variables and obtains

$$\psi_1(x,y) = \phi_1(x)e^{ik_y y}, \quad \psi_2(x,y) = ie^{-i\tau\alpha}\phi_2(x)e^{ik_y y}, \quad (3)$$

where $k_y$ is the wave vector and $\phi_{1,2}(x)$ satisfy

$$\begin{cases} v_F \hbar \phi_2' + \tau v_F(eBx + \hbar k_y)\phi_2 + (eEx - \varepsilon)\phi_1 = 0 \\ v_F \hbar \phi_1' - \tau v_F(eBx + \hbar k_y)\phi_1 - (eEx - \varepsilon)\phi_2 = 0 \end{cases}. \quad (4)$$

Since $\phi_{1,2}$ can be easily obtained for a pure magnetic field with $E=0$ as eigen-functions of a harmonic oscillator,[1] for $E \neq 0$ one may apply transforms

$$x = C\xi + D, \quad \phi_{1,2}(x) = \chi_{1,2}(\xi)e^{-\xi^2/2} \quad (5)$$

with $C$, $D$ real constants and $\chi_{1,2}$ functions to be determined. Suppose

$$L_B = \sqrt{\hbar/eB}, \quad \varepsilon_B = v_F\sqrt{\hbar eB} \quad (6)$$

respectively denote magnetic length and the energy quantum. By expanding $\chi_{1,2}$ into power series of $\xi$ one finds from Eq. (4) that for physically possible solutions $\chi_{1,2}$ must be polynomials and $C = (1-\beta^2)^{-1/4}L_B$, $D = -L_B^2 k_y \mp \tau\beta(1-\beta^2)^{-1/4}\sqrt{2n}L_B$. After determining coefficients of $\chi_{1,2}$ one obtains eigen-energies and eigen-states[5,6]

$$\varepsilon = \pm\tau(1-\beta^2)^{3/4}\sqrt{2n}\,\varepsilon_B - \beta v_F \hbar k_y, \quad (7)$$

$$\begin{pmatrix} \psi_1 \\ \psi_2 \end{pmatrix} = \begin{pmatrix} \mp(\sqrt{1-\beta^2}+\tau)\sqrt{2n}H_{n-1}(\xi) + \beta H_n(\xi) \\ ie^{-i\tau\alpha}[\pm\beta\sqrt{2n}H_{n-1}(\xi) - (\sqrt{1-\beta^2}+\tau)H_n(\xi)] \end{pmatrix} e^{-\xi^2/2+ik_y y}, \quad (8)$$

$$\xi = (1-\beta^2)^{1/4}L_B^{-1}[x + L_B^2 k_y \pm \tau\beta(1-\beta^2)^{-1/4}\sqrt{2n}\,L_B], \quad (9)$$

where $n = 0,1,2,\cdots$, $H_n(\xi) = (-1)^n e^{\xi^2} d^n e^{-\xi^2}/d\xi^n$ is the $n$ th order Hermite polynomial, and $H_{-1}(\xi) = 0$.

Solutions (7) and (8) are for $\beta < 1$ only. This limitation originates form the factor $e^{-\xi^2/2}$. When $\beta \to 1$ one has $\xi \to \pm\tau\sqrt{2n}$ and $e^{-\xi^2/2} \to e^{-n}$. Eigen-states thus become more and more extended in $x$-direction. Therefore for $\beta \geq 1$, one may adopt another two transforms

$$x = C\xi + D, \quad \phi_{1,2}(x) = \chi_{1,2}(\xi)e^{\pm i\xi^2/2} \tag{10}$$

with $C$, $D$ two another constants and $\chi_{1,2}$ two another functions. By expanding $\chi_{1,2}$ into power series of $\xi$ one finds from Eq. (4) that for physically possible solutions, $\chi_{1,2}$ must be constants and $C = (\beta^2 - 1)^{-1/4} L_B$, $D = -L_B^2 k_y$. The eigen-energies and eigen-states are

$$\varepsilon = -\beta v_F \hbar k_y, \tag{11}$$

$$\begin{pmatrix} \psi_1 \\ \psi_2 \end{pmatrix} = \begin{pmatrix} 1 \\ (\mp\sqrt{1-\beta^{-2}} - i\tau\beta^{-1})e^{-i\tau\alpha} \end{pmatrix} e^{\pm i\xi^2/2 + ik_y y}, \tag{12}$$

$$\xi = (\beta^2 - 1)^{1/4} L_B^{-1}(x + L_B^2 k_y). \tag{13}$$

Similarly, for a pure electric field with $B = 0$, one finds from Eq. (4) $C = \sqrt{v_F \hbar / eE}$, $D = \varepsilon / eE$, and $k_y = 0$. The eigen-energy may take any real value and the eigen-states are

$$\begin{pmatrix} \psi_1 \\ \psi_2 \end{pmatrix} = \begin{pmatrix} 1 \\ \mp e^{-i\tau\alpha} \end{pmatrix} e^{\pm i\xi^2/2}, \tag{14}$$

$$\xi = \sqrt{eE / \hbar v_F}(x - \varepsilon / eE). \tag{15}$$

Different from the case of $\beta < 1$, for $\beta > 1$ or a pure electric field, the two states in Eq. (12) or (14) distinguished by different signs have the same eigen-energy. Equation (12) or (14) thus presents two branches of eigen-states and the degeneracy will be called branch degeneracy. This is the third degeneracy in addition to spin degeneracy and valley degeneracy. It can be proven that states in Eq. (12) and those in Eq. (14) are connected by Lorentz transform for $\beta > 1$.[17]

The expectation value of velocity operator $\hat{\vec{v}} = v_F \vec{\sigma}$ for a non-normalized eigen-state can be calculated by

$$\vec{v} = \lim_{\substack{a \to +\infty \\ b \to +\infty}} \frac{v_F \int_{-b}^{+b} dy \int_{-a}^{+a} (\psi_1^* \ \psi_2^*) \vec{\sigma} (\psi_1 \ \psi_2)^T dx}{\int_{-b}^{+b} dy \int_{-a}^{+a} (\psi_1^* \ \psi_2^*)(\psi_1 \ \psi_2)^T dx}. \tag{16}$$

For $\beta < 1$, $\beta \geq 1$, and a pure electric field, Eqs. (8), (12), and (14) respectively lead to

$$\vec{v} = -\beta v_F \vec{y} \quad (17), \qquad \vec{v} = \mp\sqrt{1-\beta^{-2}}\, v_F \vec{x} - \beta^{-1} v_F \vec{y} \quad (18), \qquad \vec{v} = \mp v_F \vec{x} \quad (19).$$

These velocities depict ballistic motion of graphene's carriers in a field.[18] For $\beta > 1$ or a pure electric field, two branches of eigen-states have different velocities.

Carriers in solids usually stay in wave-packet states.[19] A Gaussian wave packet of graphene is formed by $(\tilde{\psi}_1 \ \tilde{\psi}_2) = C_0 \int_{-\infty}^{+\infty} (\psi_1 \ \psi_2) \exp[-(\varepsilon - \varepsilon_0)^2 / \Delta_\varepsilon^2 - i\varepsilon(t-t_0)/\hbar] d\varepsilon$, where $C_0$, $t_0$ are constants, $\varepsilon_0$ is the central energy, and $\Delta_\varepsilon$ is the half-width of the energy range. The probability density is given by $\rho = |\tilde{\psi}_1|^2 + |\tilde{\psi}_2|^2$. For $\beta < 1$, Eq. (8) leads to

$$\rho(x, y, t) = P(x) \exp\left[ -\frac{(x-x_0)^2}{\Delta_x^2} - \frac{(y-y_0)^2}{\Delta_y^2} \right], \tag{20}$$

$$x_0 = -L_B^2 k_{y,0} \mp \tau\beta(1-\beta^2)^{-1/4}\sqrt{2n}\, L_B, \quad y_0 = -\beta v_F(t-t_0), \tag{21}$$

$$\Delta_x = L_B \sqrt{(1-\beta^2)^{-1/2} + \frac{\Delta_\varepsilon^2}{2\beta^2 \varepsilon_B^2}}, \quad \Delta_y = L_B \sqrt{(1-\beta^2)^{1/2} + \frac{2\beta^2 \varepsilon_B^2}{\Delta_\varepsilon^2}}, \tag{22}$$

where $P(x)$ is a polynomial and $k_{y,0}$ is the wave vector of the central state. Both $\Delta_x$ and $\Delta_y$ are independent of time, and Eq. (20) describes a 2D soliton that propagates along $y$-axis at the velocity given by Eq. (17). For $\beta \geq 1$, Eq. (12) leads to

$$\rho(x,y,t) = \frac{1}{\sqrt{\pi}\Delta} \exp\left[-\frac{(\pm\sqrt{1-\beta^{-2}}x + \beta^{-1}y + v_F t - d)^2}{\Delta^2}\right], \tag{23}$$

$$d = \pm\sqrt{\beta^2 - 1}\,\varepsilon_0 L_B / \beta^2 \varepsilon_B + v_F t_0, \tag{24}$$

$$\Delta = L_B \sqrt{\frac{(\beta^2-1)\Delta_\varepsilon^2}{2\beta^4 \varepsilon_B^2} + \frac{2\varepsilon_B^2}{\Delta_\varepsilon^2}}. \tag{25}$$

Equation (23) describes an 1D soliton that propagates at the velocity given by Eq. (18) and whose peak forms into a ridge along the straight line $\pm\sqrt{1-\beta^{-2}}\,x + \beta^{-1}y + v_F t - d = 0$. For a pure electric field, Eq. (14) leads to

$$\rho(x,t) = \frac{1}{\sqrt{\pi}\Delta_E} \exp\left[-\frac{(x \pm v_F t - d_E)^2}{\Delta_E^2}\right], \tag{26}$$

$$d_E = \varepsilon_0 / eE \pm v_F t_0, \tag{27}$$

$$\Delta_E = \sqrt{\frac{\Delta_\varepsilon^2}{2e^2 E^2} + \frac{2v_F^2 \hbar^2}{\Delta_\varepsilon^2}}. \tag{28}$$

Equation (26) describes an 1D soliton that propagates along the electric field at the velocity given by Eq. (19). For $\beta \geq 1$ and a pure electric field, the two branches of wave packets distinguished by different signs correspond to two branches of eigen-states.

Eigen-states for the whole range of $\beta$ provide a complete picture of their evolution with the field and LL collapse. When $\beta$ increases form 0 to 1, eigen-states with a fixed $k_y$ and different $n$ evolve continuously from different states of $\beta=0$ to the same state of $\beta=1$ with the same $k_y$, and when $\beta$ continues to increase form 1, this state evolves into two degenerate branches. For a fixed $E$, when $\beta \to +\infty$ all states in the same branch with different $k_y$ tend to the same state of the pure electric field. The complete eigen-states also provide a complete picture of graphene's carrier motion in different fields: As $\beta$ increases from 0 to 1, the carrier's velocity increases its magnitude from 0 to $v_F$ but remains in the direction of $\vec{E} \times \vec{B}$. As $\beta$ continues to increase from 1 to $+\infty$, the velocity maintains its magnitude at $v_F$ and gradually turns to the electric field, either right-handwise or left-handwise according to different branches. Solitons are 2D for $0 < \beta < 1$ and become 1D for $\beta \geq 1$. Since $\Delta_x$, $\Delta_y$ given by Eq. (22) are steep functions as $\beta \to 1$, solitons go through an abrupt dimensional transition at $\beta=1$.

In addition to eigen-states themselves, the density of states plays an important role in electronic transport. Suppose the graphene has length $L_x$ in $x$-direction and width $L_y$ in $y$-direction. For $\beta<1$, the number of states $N$ for each LL is determined according to the periodical boundary condition in $y$-direction and the confinement of states to graphene in $x$-direction.[20,21] With spin degeneracy and valley degeneracy not included, one has

$$N = L_x L_y / 2\pi L_B^2 = eB L_x L_y / h. \tag{29}$$

This number is just that of magnetic-flux quanta $h/e$ in the graphene,[5,22] since each quantum has the area $S_B = h/eB$. It can be proven that this number also equals the maximum number of Gaussian wave packets of Eq. (20). According to Eq. (22), $\Delta_x$, $\Delta_y$ take their minimum values $\Delta_{x,\min}$, $\Delta_{y,\min}$ at $\Delta_\varepsilon = \sqrt{2}\beta(1-\beta^2)^{-1/4}\varepsilon_B$, and

$$\Delta_{x,\min} = \sqrt{2}(1-\beta^2)^{-1/4}L_B, \quad \Delta_{y,\min} = \sqrt{2}(1-\beta^2)^{1/4}L_B. \tag{30}$$

Each wave packet can be regarded as occupying an elliptic area given by $(x-x_0)^2/\Delta_{x,\min}^2 + (y-y_0)^2/\Delta_{y,\min}^2 \leq 1$, and thus has the area $\pi\Delta_{x,\min}\Delta_{y,\min} = h/eB = S_B$. The maximum number of wave packets is thus given by Eq. (29).

For $\beta > 1$ one may determine the number of states by calculating the maximum number of Gaussian wave packets. According to Eq. (25), $\Delta$ takes its minimum value $\Delta_{\min}$ at $\Delta_\varepsilon$, with

$$\Delta_{\min} = \sqrt{2}\beta^{-1}(\beta^2-1)^{1/4}L_B, \quad \Delta_\varepsilon = \sqrt{2}\beta(\beta^2-1)^{-1/4}\varepsilon_B. \tag{31}$$

According to Eq. (18), the projective length of the graphene's diagonal in the direction of wave-packet velocity is

$$L_d = \sqrt{1-\beta^{-2}}L_x + \beta^{-1}L_y. \tag{32}$$

With spin degeneracy and valley degeneracy not included, the maximum number of wave packets for each branch is

$$N = L_d/2\Delta_{\min}. \tag{33}$$

When the graphene is filled up with wave packets, at any time centers of two neighboring wave packets in the same branch have distance $2\Delta_{\min}$. According to Eq. (24), the central-energy gap of the neighboring wave packets is just the width of energy range $2\Delta_\varepsilon$ with $\Delta_\varepsilon$ given by Eq. (31). One concludes that near graphene's neutral point where $\varepsilon = 0$, energy levels are

$$\varepsilon = \pm 2\sqrt{2}\beta(\beta^2 - 1)^{-1/4} n\varepsilon_B, \tag{34}$$

where $n = 0, 1, 2, \cdots$, and $+$, $-$ in $\pm$ are respectively for electrons and holes.

Similarly, for a pure electric field, $\Delta_E$ in Eq. (28) takes its minimum value $\Delta_{E,\min}$ at $\Delta_{\varepsilon,E}$, with

$$\Delta_{E,\min} = \sqrt{2v_F\hbar/eE}, \quad \Delta_{\varepsilon,E} = \sqrt{2v_F\hbar eE}. \tag{35}$$

With all three kinds of degeneracy not includes, the maximum number of wave packets is

$$N = L_x / 2\Delta_{E,\min}. \tag{36}$$

According to Eq. (27), the central-energy gap of two neighboring wave packets in the same branch is just the width of energy range $2\Delta_{\varepsilon,E}$ with $\Delta_{\varepsilon,E}$ given by Eq. (35). Near $\varepsilon = 0$ the energy levels are

$$\varepsilon = \pm 2n\sqrt{2v_F\hbar eE}, \tag{37}$$

where $n = 0, 1, 2, \cdots$, and $+$, $-$ in $\pm$ are respectively for electrons and holes.

In the case of Hall effect, the source and drain electrodes are respectively connected to the lower and upper edges of the graphene. The current is in $y$-direction and the voltage $V$ in $x$-direction is the Hall voltage which is either generated by the accumulated charge at the left and right edges or applied externally. For $\beta < 1$, Hall conductance can be obtained according to eigen-states (8) and their velocity (17).[20,21]

For $\beta > 1$, carriers with $\vec{v} = -\sqrt{1-\beta^{-2}} v_F \vec{x} - \beta^{-1} v_F \vec{y}$ and those with $\vec{v} = +\sqrt{1-\beta^{-2}} v_F \vec{x} - \beta^{-1} v_F \vec{y}$ have the same velocity component $v_y = -\beta^{-1} v_F$. Two kinds of carriers are exchanged by reflection at the left or right edge, and therefore have the same Fermi energy $\varepsilon_F$. Since each electron or hole respectively contributes to the current by $\pm e v_y / L_y$, at temperature $T$ the current can be calculated by

$$I_y = \frac{8 e v_F}{\beta L_y} \left\{ \sum_{n=0}^{+\infty} c_n f(|\varepsilon|-\varepsilon_F) - \sum_{n=0}^{+\infty} c_n [1 - f(-|\varepsilon|-\varepsilon_F)] \right\}, \tag{38}$$

where $c_0 = 1/2$, $c_n = 1$ for $n \neq 0$, $\varepsilon$ is given by Eq. (34), $f(\varepsilon) = 1/(1+e^{\varepsilon/k_B T})$ is the Fermi-Dirac distribution function with $k_B$ the Boltzmann constant, the factor 8 comes from the three kinds of degeneracy, and the two sums respectively represent contributions of electrons and holes. Since $\beta = V / v_F B L_x$, for a fixed $B$, $\varepsilon$ is a function of $V$ and Eq. (38) presents the relation between the current and the Hall voltage. Because $\beta > 1$, the voltage must satisfy $V > v_F B L_x$. The current depends on Hall voltage nonlinearly. For large $V$, approximately $I_y \propto 1/V$ and this is the effect of the factor $\beta^{-1}$ in velocity (18). Numerical results are presented in Fig. 1.

In the case of a pure electric field, the source and drain electrodes are respectively connected to the left and right edges of the graphene, and the voltage $V$ is applied to these electrodes. With reflection at the electrodes neglected, carriers with $\vec{v} = +v_F \vec{x}$ come from the left electrode and have Fermi energy $\varepsilon_F$ which is also the chemical potential of the left electrode, and carriers with $\vec{v} = -v_F \vec{x}$ come from the right electrode and have Fermi energy $\varepsilon_F + eV$ which is also the chemical potential of the right electrode.[23] The current is in $x$-direction and can be calculated by

$$I_x = -\frac{4ev_F}{L_x}\left\{\sum_{n=1}^{+\infty}f(|\varepsilon|-\varepsilon_F) - \sum_{n=1}^{+\infty}f(|\varepsilon|-\varepsilon_F-eV)\right.$$
$$\left. -\sum_{n=1}^{+\infty}[1-f(-|\varepsilon|-\varepsilon_F)] + \sum_{n=1}^{+\infty}[1-f(-|\varepsilon|-\varepsilon_F-eV)]\right\}, \quad (39)$$

where $\varepsilon$ is given by Eq. (37), the first two sums represent the contribution of electrons and the last two represent that of holes, and the factor 4 comes from spin degeneracy and valley degeneracy. Since in Eq. (37) $E = V/L_x$, Eq. (39) presents the current-voltage relation.

When $L_x = 2n\Delta_{E,\min}$ with $n = 1, 2, \cdots$, the graphene has at most $4n$ carriers. According to Eq. (35), corresponding voltage is

$$V_n = 8n^2 v_F \hbar / eL_x. \quad (40)$$

At zero temperature, for $V_n \leq V < V_{n+1}$ the current is

$$I_{x,n} = 4nev_F / L_x. \quad (41)$$

When $V < V_1$, there will be no current and the threshold voltage is given by $V_1 = 8v_F\hbar/eL_x$. The current thus demonstrates steps at low temperature and reflects the effect of only four carriers. This is different from the case of $\beta < 1$, where the number of states for each LL is much larger. For large $V$, approximately $I_x \propto \sqrt{V}$. This is because according to Eq. (36), the number of states $N \propto \sqrt{V}$. Numerical results are presented in Fig. 2.

Gaussian wave packets provide a way of understanding carrier motion of graphene in an electromagnetic field. In this picture, carriers behave as solitons propagating in fixed velocities, and two neighboring solitons are separated by the minimum width, reflecting Pauli exclusion principle. Solitons thus resemble particles. Nevertheless, each soliton accommodates four carriers, demonstrating identical motion of carriers with different spin or $\vec{k}_F$. Solitons propagating at different velocities can pass through each other without being affected. Besides, an 1D soliton is still extended in the direction perpendicular to its velocity. Therefore solitons are different from particles. It should also be noted that different wave packets can be constructed depending on different superposition.[24] For $\beta < 1$, the uniform motion of wave packets in this work is different from the cycloid-like one obtained by quasi-classical approach.[21,25] This work thus presents a simpler picture but enriches the mystery of wave-particle duality of carriers in graphene. In addition, the maximum number of Gaussian wave packets provides a novel approach to the determination of the number of states. In fact, periodical boundary condition is just a principle adopted without any proof.[19] Alternatively, solitons are arranged closely one by one to occupy the entire graphene area, and this presents a reasonable explanation to the number of states. Finally, the step-like current-voltage curves of graphene in a pure electric field demonstrate the phenomenon similar to the single-electron effect. However, the material here is novel and different features regarding the height of the current steps and the length of the voltage plateaus may be respected. In conclusion, eigen-states of graphene in an electromagnetic field and their solitons may be helpful to the investigations of some fundamental problems of quantum mechanics and few-electron physics.

This work was supported by the NSF EPSCOR (grants 1002410 and 1010094).

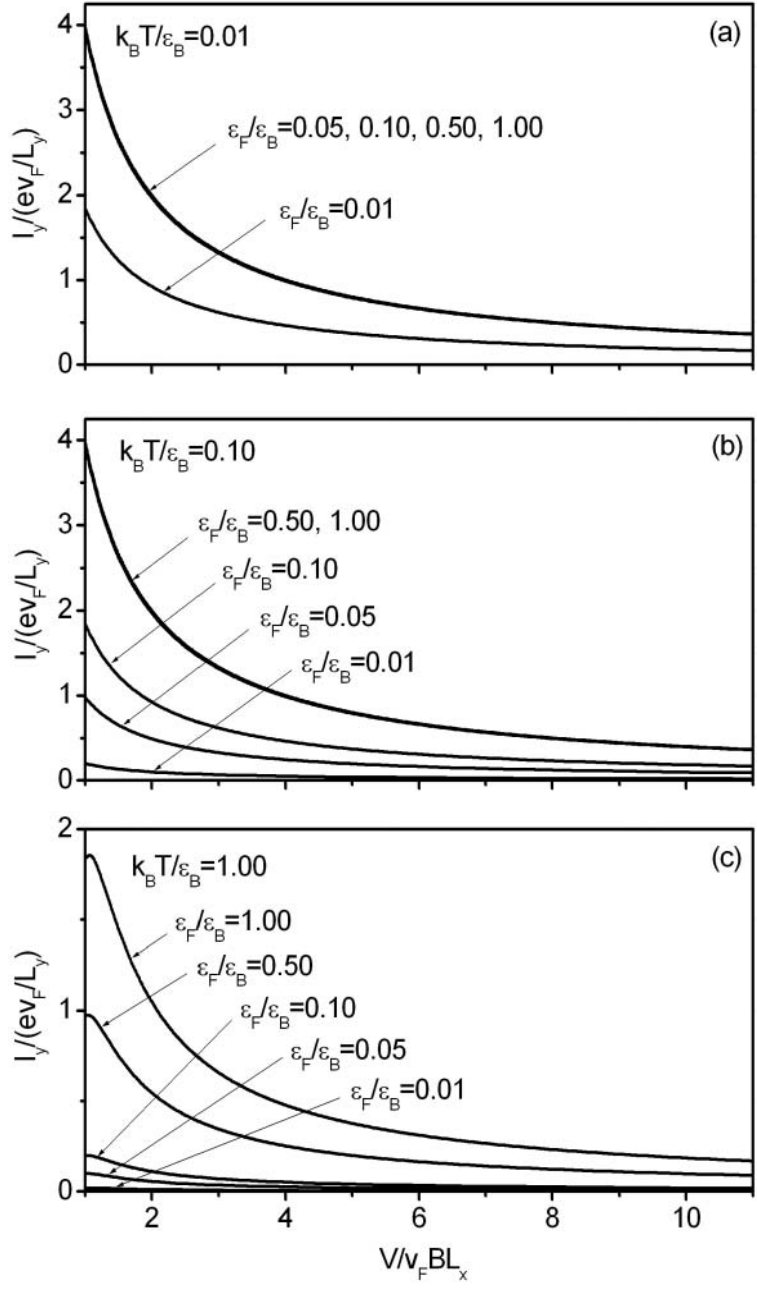

FIG. 1. Normalized current $I_y / ev_F L_y^{-1}$ as a function of normalized Hall voltage $V / v_F B L_x$ of graphene in a magnetic field for $V > v_F B L_x$, calculated according to Eq. (38) at different normalized temperatures $k_B T / \varepsilon_B$ and for different normalized Fermi energies $\varepsilon_F / \varepsilon_B$, with $\varepsilon_B$ given by Eq. (6). Curves are symmetric with respect to the origin.

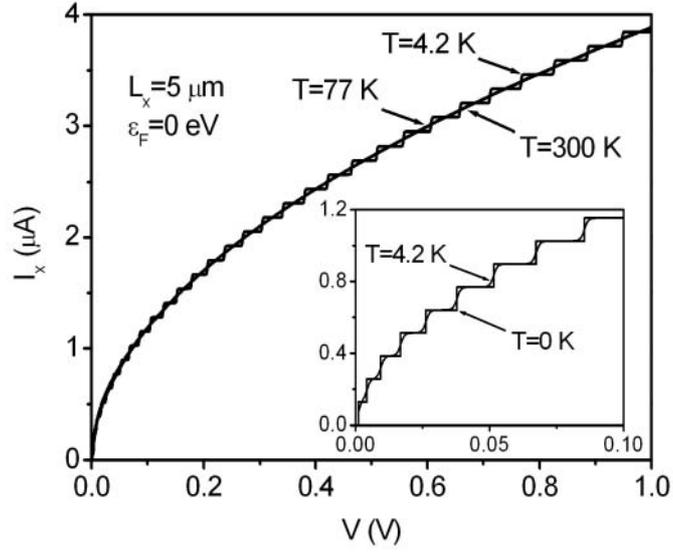

FIG. 2. Current-voltage curves of graphene without a magnetic field, calculated according to Eq. (39) at different temperatures. Inset: Current steps at low temperatures, including those at $T = 0$ K determined by Eq. (41). Curves are symmetric with respect to the origin.